# Magneto-Optical Cooling of Atoms

Mark G. Raizen[1], Dmitry Budker[2,3], Simon Rochester[3],
Julia Narevicius[4], and Edvardas Narevicius[4]

[1] Department of Physics, University of Texas at Austin, Austin, TX 78712, USA

[2] Department of Physics, University of California at Berkeley, Berkeley, CA 94720, USA

[3] Rochester Scientific, El Cerrito, CA 94530, USA (www.rochesterscientific.com)

[4] Dept. of Chemical Physics, Weizmann Institute of Science, Rehovot, Israel

## Abstract

We propose an alternative method to laser cooling. Our approach utilizes the extreme brightness of a supersonic atomic beam, and the adiabatic atomic coilgun to slow atoms in the beam or to bring them to rest. We show how internal-state optical pumping and stimulated optical transitions, combined with magnetic forces can be used to cool the translational motion of atoms. This approach does not rely on momentum transfer from photons to atoms, as in laser cooling. We predict that our method can surpass laser cooling in terms of flux of ultra-cold atoms and phase-space density, with lower required laser power and reduced complexity.

Over the past thirty-eight years since its invention [1,2], laser cooling has remained as the standard approach for producing ultra-cold atoms [3]. This method relies on the momentum transfer from light to atoms, as photons are repeatedly scattered, enabling the study of ultra-cold atomic gases. While laser cooling works very well, the requirement of a closed, two-level transition has limited the applicability of the method to a small subset of elements in the periodic table. For those elements, laser cooling has reached saturation in the number of ultra-cold atoms produced per second, atomic density, phase-space density, and number of photons per atom cooled. A natural question is whether one can find an alternative approach to laser cooling that can break the barriers of generality and performance.

In a separate development over the past fifty years, physical chemists invented and refined ultra-bright sources of atoms and molecules known as supersonic

(molecular) beams [4]. These beams are created by the expansion of a dense carrier gas from a small aperture, and are typically operated in a pulsed mode. The supersonic beam serves as a universal source of fast but cold atoms or molecules, which are mixed or entrained in the carrier-gas flow. The mixture cools as it expands, and becomes collision-less downstream from the nozzle. In previous work, paramagnetic species were decelerated and stopped with an atomic and molecular coilgun, which utilizes pulsed magnetic fields [5]. In a more recent development, an adiabatic coilgun was shown to be well-matched to the beam, translating it to the desired terminal velocity with almost no loss of phase-space density [6,7]. Independently, cooling of magnetically trapped atoms was demonstrated using so-called single-photon cooling. This method does not rely on the momentum of the photon, but rather on a one-way wall, realizing the historic thought experiment of Maxwell's Demon [5]. Single-photon cooling achieved a 350x enhancement of phase-space density, so proved the concept experimentally. However, it does not seem practical to apply this method to a magnetically stopped supersonic beam because the capture efficiency would be too low. Until now, it was not clear how to optimally integrate the above techniques, and a direct comparison with laser cooling was therefore not possible.

In this Letter, we propose a new approach; magneto-optical (MOP) cooling in six-dimensional phase space. It relies on internal-state optical pumping and stimulated optical transitions, combined with magnetic forces to reduce the translational motion of atoms. This approach does not rely on momentum transfer from photons to atoms, as in laser cooling, so is fundamentally different and new. The method works independently on each of the six degrees of freedom in phase space, unlike a trap where they are coupled. We analyze the proposed method, and predict that the performance can be far superior to laser cooling. Our quantitative predictions are supported by realistic numerical simulations.

The starting point is a pulsed supersonic beam, consisting of a pulsed valve and a skimmer contained in a region that is differentially pumped. The current state-of-the-art valve is the Even-Lavie design, which produces beams with a brightness of over $10^{22}$ atoms/sr/s, in pulses of 20 μs FWHM, and at a repetition rate of up to 1 kHz [8]. Each pulse of the Even-Lavie valve contains around $10^{16}$ carrier-gas atoms. Atoms or molecules can be entrained into the carrier gas, and negligible heating has been reported for fractional entrainment of 1% or less [9,10]. Supersonic beams with atom densities of $10^{13}$ cm$^{-3}$ in the interaction region (past the skimmer) have been created by laser ablation [9]. There are about $10^{14}$ entrained atoms in each pulse, and of those only 3% get through the skimmer, giving $3 \times 10^{12}$ atoms. The atoms in the beam can be brought to rest in the laboratory frame, and adiabatically released to free space. We now describe how the 6-D phase space can be efficiently compressed, resulting in an ultra-cold and dense cloud of unprecedented brightness.

We first consider atoms with no hyperfine structure, and assume a J=1 ground state, and J'=1 excited state to be specific. A realistic example used for our simulations is metastable $^4$He, one of the standard atoms amenable to laser cooling. We first consider one dimension, denoted as x, which can then be extended to 3-D. A schematic illustrating the basic method is shown in Fig. 1(a,b). Compression of velocity in the x dimension, $v_x$, is shown in Fig. 1(a), while spatial compression in the x dimension is shown in Fig. 1(b). There are three basic tools that are required for this cooling and spatial compression: magnetic-state optical pumping, velocity-selective stimulated rapid adiabatic passage (STIRAP), and one-dimensional magnetic kicking. Optical pumping is the most standard of these three tools, dating back to the work of A. Kastler, and reviewed in Ref. [11]. The method of STIRAP enables stimulated transitions between internal states of atoms or molecules, and is relatively insensitive to variations in the laser parameters, such as intensity, pulse duration, and pulse overlap [12]. Due to its inherent robustness, STIRAP has become a widely used method in atomic and molecular physics. A more recent twist on STIRAP is that it can be applied in a geometry that is velocity-selective, as proposed in Ref. [13].

The last tool, one-dimensional magnetic kicking, is perhaps least known, although the basic idea is a variation of the historic Stern-Gerlach effect. An atom with a magnetic moment $\boldsymbol{\mu}$ subject to a non-uniform magnetic field $\mathbf{B}$ experiences a gradient force $\mathbf{F} = \nabla(\boldsymbol{\mu} \cdot \mathbf{B})$. Thus, for example, the x component of the velocity of a subgroup of atoms in the m=1 state (i.e., with magnetic moment along the z direction) can be manipulated using a magnetic field that points in the z direction and varies along the x axis. Due to the condition $\nabla \times \mathbf{B} = 0$ for magnetic fields, such a field will also have a gradient in a transverse direction. Thus one might expect undesirable transverse momentum spreading, as well as spin mixing, induced by the transverse gradient. A similar situation arises when considering a magnetic field with a gradient along the field direction, due to the divergence condition $\nabla \cdot \mathbf{B} = 0$. However, as noted in the proposal for the original Stern-Gerlach experiment, the application of a bias magnetic field along the z direction causes precession about the z axis, effectively averaging the transverse forces to zero [14,15]. In the Supplementary Material we verify with a detailed quantum mechanical calculation that as long as the bias field is large compared with the product of the transverse field gradient and the spatial extent of the atomic cloud, the transverse kicks and spin mixing are effectively eliminated. Thus, with an appropriate field configuration, including a sufficiently large bias field, an ensemble of atoms in the m=1 state with respect to any given quantization axis can be given a 1-D momentum kick transverse to that axis.

    We now describe in detail the application of these three tools in the cooling and spatial compression sequences. The first step in Fig. 1(a) is to optically pump all of the atoms into the m=0 ground state, using linearly polarized light. A two-photon STIRAP pulse is used to drive the transition from m=0 to m=1 for a subset of the atoms with a non-zero mean velocity in the x dimension. This can be accomplished, for example, by using beams counter-propagating along the x axis, one with linear polarization along z and the other with linear polarization along y, with a z-directed magnetic bias field to select the transition driven by the left-circularly polarized component of the y-polarized light. Now that atoms in a velocity class are in the m=1 state, a pulsed, uniform magnetic field directed along z, with a gradient along x is applied to impart momentum to these atoms in order to shift the mean velocity to a desired final velocity. A resonant pulse is then applied to optically pump these atoms back to m=0, using linearly polarized light. The other atoms in the m=0 state will not be affected because the m=0 to m'=0 transition is forbidden. The cycle is then repeated for other velocity classes, until all atoms are concentrated around a desired final velocity in the laboratory frame. One realization of the sequence is to *halve* the width of the velocity distribution in each

cycle. This would enable geometrically-fast compression of velocity space, as illustrated in Fig. 1 (a). For example, a six-step process would compress the velocity spread by $2^6$, and each atom must be optically pumped on average three times. Since the process is Doppler sensitive, it can be applied sequentially in 3-D. We estimate that the cooling limit of our method is comparable to *sub-Doppler* laser cooling, due to the random-walk in momentum from spontaneously scattered photons in the optical-pumping stages. Note that the compression process does not have to produce a thermal distribution, nor does it have to be symmetrical in velocity space. We can estimate the required laser power for the entire cooling process. Each cycle requires a STIRAP pulse, and the maximum power is required for the first step, where the initial velocity distribution is halved. The typical velocity spread of the stopped atoms prior to cooling is around 10 m/s, so the STIRAP pulse must cover a velocity class of around 5 m/s. We simulated this process numerically, and find that it works as predicted. Representative parameters are an intensity of 90 mW/cm$^2$ for each of the two STIRAP beams, detuned 30 MHz from resonance, and pulse durations of 200 ns. After this cooling cycle, further compression in momentum near or even below the recoil limit can be accomplished with stimulated Raman cooling [16,17].

This velocity compression is reminiscent of the compression of the density of particles in a box, as in the original Maxwell Demon thought experiment, although here the box is in momentum space rather than positional space. The resonant laser is used for optical pumping, while the magnetic field gradient does the work. In fact, it works much like a mop, sweeping the atoms around and collecting them into small piles, hence the acronym MOP is quite appropriate. Unlike the original version of Maxwell's Demon, the atoms are free particles and will expand ballistically, as they are not contained by any boundaries. This means that the cooling cycles must be fast enough so that free-space expansion is insignificant. The only irreversible step in the cycle is the optical pumping stage, which carries away the entropy, saving the Second Law of Thermodynamics.

We now propose how to compress the volume occupied by the atoms without heating, unlike adiabatic compression. An optical pumping beam is made to spatially overlap one half of the atoms, pumping them from the m=0 to the m=1 state, as shown in Fig.1 (b). A uniform magnetic field gradient is then applied which first accelerates the atoms, then decelerates them, so that in the end they have the same velocity profile, but have moved over to the location of the other half of the atoms. Finally, these atoms are optically pumped back to m=0. At the end of this cycle, the density is increased by a factor of 2. This spatial compression can be applied sequentially in all three dimensions. It is natural to inquire about the density limit of this process. The main limitation is that at sufficiently high density, photon re-scattering by adjacent atoms will degrade the fidelity of optical pumping [18]. This effect is predicted to occur at a density of $10^{11}$-$10^{12}$ atoms/cm$^3$, significantly higher than the density limit of laser cooling (around $10^9$ atoms/cm$^3$). The high density can also be converted to lower temperature by the following approach: after spatial compression in 3-D, a magnetic trap is turned on suddenly. In order to minimize heating, the trap depth is chosen to match the distributions in position and momentum. The trap is then released adiabatically, lowering the temperature at the expense of density. The process can be repeated until the momentum distribution is minimized, and finally, the spatial distribution is compressed to maximum density.

The same process works as well for atoms with hyperfine structure, and with an odd nuclear spin where there is no m=0 state in strong magnetic fields. A realistic example is $^7$Li, where the ground state hyperfine splitting is 803 MHz. Optical pumping requires a re-pump beam that is offset by the hyperfine splitting. In moderate magnetic fields on the order of 100 G, the electronic spin decouples from the nuclear spin, so that one can assume a J=1/2 atom without nuclear spin to estimate the magnetic forces. The same sequence of cooling and compression works in this case, but the optical pumping is between m=-1/2 and m=1/2 states. The magnetic field gradient then acts in a push-pull configuration. Similarly, compression in real space can be accomplished by switching the direction of the field gradient.

We now estimate the number of atoms per second, and the phase-space density that can be expected. MOP cooling will compress the volume in 3D velocity space, and spatial compression in 3D real space will further increase phase-space density. The

Even-Lavie valve can operate at a repetition rate of up to 1 kHz, implying 3x $10^{15}$ ultra-cold atoms per second, a factor of around $10^6$ more than reached with laser cooling. The phase-space density of the supersonic beam after the skimmer is around $10^{-8}$, so after compression it should be near $10^{-3}$, a factor of $10^3$ higher than in laser cooling. The small number of spontaneous emission events per atom implies that many elements require only a single-color laser, even if they do not have closed cycling transitions.

We have performed 3D Monte-Carlo simulations of the proposed scheme. We assume a cloud of atoms initially at rest with standard velocity deviation in all the coordinates equal to 10 m/s. The first two key steps in the cooling sequence are shown in Fig. 2 (a-f). The initial phase-space and velocity distributions of 10,000 atoms along the x-coordinate are presented in Fig. 2 (a) and (b) respectively. Figures 2 (c) and (d) illustrate the result of velocity-selective STIRAP population transfer of atoms from m=0 to m=1 state, atoms in m=1 state are represented by red dots. The atoms are selected with velocities centered at -10 m/s with standard velocity deviation of 5 m/s. The selected velocity group is shifted to a mean velocity of 0 m/s by applying a 60 µs-long constant-gradient magnetic field. The duration of a magnetic gradient "kick" is long enough to decelerate metastable helium atoms with 10 m/s initial velocity assuming a 60 T/m magnetic field gradient. This sequence is repeated 36 times, overall, alternating between different coordinates and magnetic gradient kick directions. Since the center velocity spread is approximately halved in each cooling cycle, we reduce the magnetic-gradient-"kick" duration accordingly. The shortest-"kick"-duration was limited to 10 µs. The final phase-space and velocity distributions are shown in Fig.3. The momentum is compressed by a factor of $10^3$ for 4000 atoms out of original 10,000. Our simulations show a compression of phase-space volume by more than two orders of magnitude. The required magnetic field gradients and short switching times are very realistic, based on demonstrated capability with the atomic coilgun [5]. A simple and practical magnetic coil configuration is a combination of Helmholtz and anti-Helmholtz circular coils.

In summary, we propose a method to produce a new source of ultra-cold atoms with unprecedented brightness. The atoms can be stationary in the laboratory frame after cooling, or they can be launched into a beam with the same magnetic-field gradients that are used for cooling. Our cooling scheme can be also applied to molecules whenever optical pumping is possible [19,20]. It is interesting to consider the impact of such a new source on science and technology. The atom laser, hailed as an enabling development, is created by launching a Bose-Einstein condensate. Starting with much higher numbers of atoms and higher phase-space density would result in a much brighter atom laser. One application is in atomic interferometry, both for inertial sensing, and for fundamental tests [21,22]. Another application where higher flux would be important is in atomic lattice clocks [23,24]. A long-standing goal has been to control atoms at the nanoscale by focusing atomic beams onto surfaces (atom lithography). This dream has not been realized so far, mainly due to limitations of lens aberrations, together with weak atomic flux. The presently proposed ultra-bright sources, together with an aberration-corrected pulsed magnetic lens [25], should enable the realization of the full potential of atom lithography. This would have important applications in materials science and condensed matter physics.

M. G. R. acknowledges support from the U.S. National Science Foundation, the R.A. Welch Foundation, Grant number F-1258, and the Sid W. Richardson Foundation. D. B. acknowledges support by the Miller Institute for Basic Research in Science. E. N. acknowledges the historic generosity of the Harold Perlman Family and support from the Israel Science Foundation.

**Figure Captions**

**Fig. 1.** Schematic of MOP cooling. (a) Momentum-space cooling in which atoms are first optically-pumped into the m=0 state. A velocity-selective STIRAP pulse is then applied to transfer all atoms in the left half of the velocity distribution into the m=1 state. A magnetic field gradient is next applied to translate the mean velocity of the selected atom ensemble to overlap with the m=0 state atoms. Optical-pumping is finally applied to drive the atoms back to the m=0 state. (b) Spatial focusing: spatially-selected atoms are first optically-pumped into the m=1 state. Magnetic field gradients are then used to translate the selected atoms towards the m=0 atoms. Optical-pumping finally drives atoms back to the m=0 state.

**Fig. 2.** Numerical simulation results showing snapshots of the phase-space distribution taken at different cooling stages. (a) and (b) show the initial phase-space distribution. (c) and (d) show the phase-space distribution after applying the STIRAP pulse. Atoms pictured in red (stars) are transferred to the m=1 state. Black circles represent the rest of the atoms that remain in the m=0 state. (e) and (f) show phase-space distributions after applying a uniform magnetic field gradient that gives a velocity kick, overlapping the atoms in velocity space. Note, that grey histograms represent velocity distributions for all of the atoms whereas red histograms show contribution from the m=1 atoms only.

**Fig. 3.** Numerical-simulation results showing the final phase-space (a) and velocity distributions (b) along one of the coordinates.

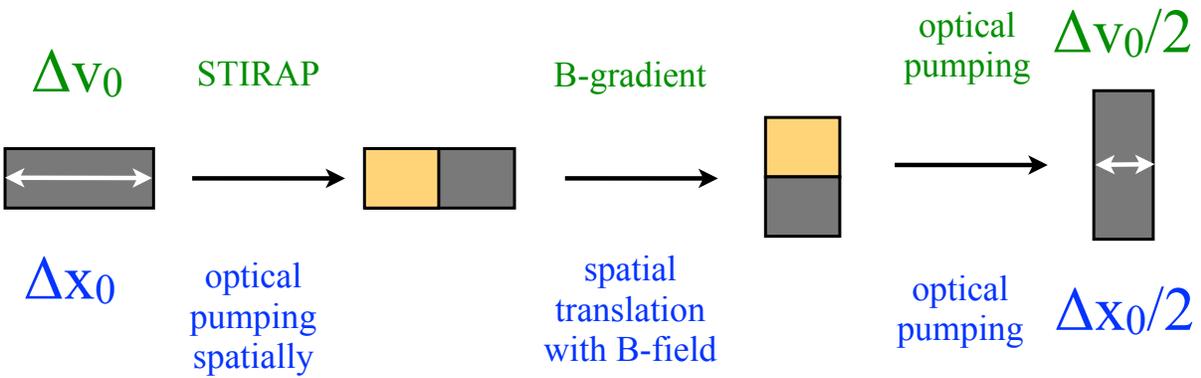

Figure. 1

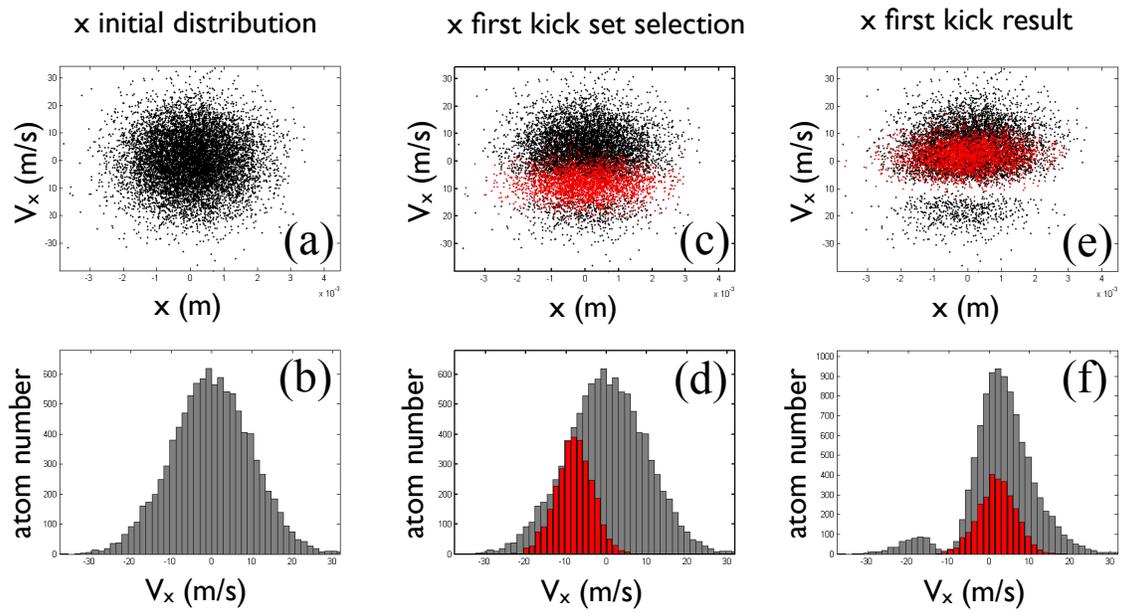

Figure. 2

## x-final distribution

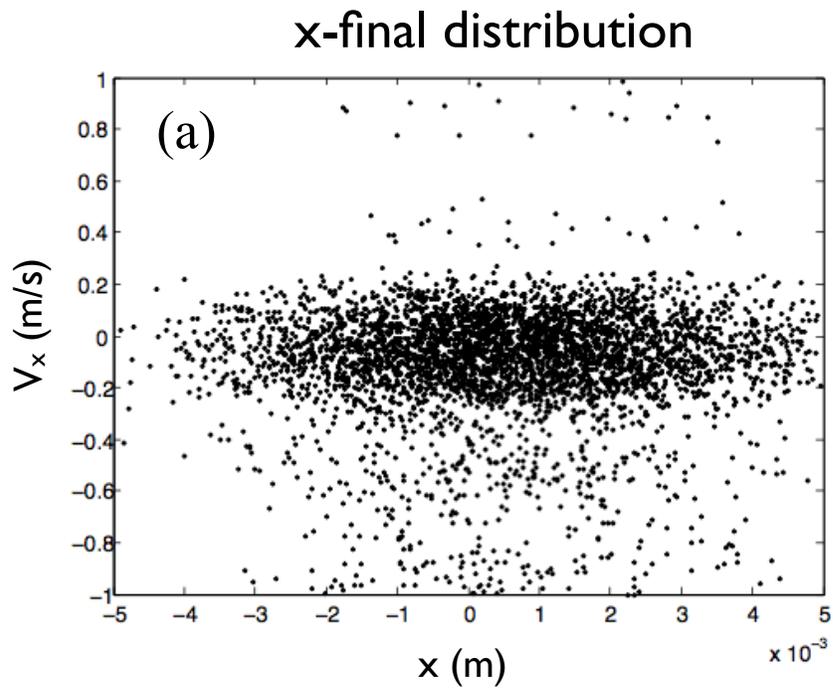

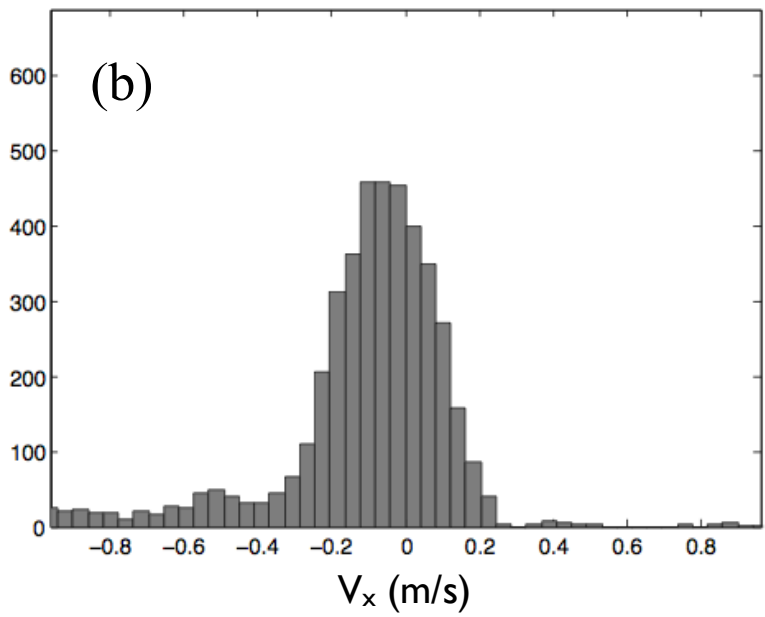

Figure. 3